\documentclass[twocolumn]{jpsj3}
\usepackage{ulem}
\usepackage{color}
\usepackage{braket}
\usepackage{amsmath}
\definecolor{mygreen}{rgb}{0.0,0.7,0.0}

\def\braket#1{\mathinner{\langle{#1}\rangle}}
\usepackage{amssymb}
\usepackage{bm}

\title{ Charge Order and Superconductivity \\ as Competing Brothers in Cuprate High-$T_{\rm c}$ Superconductors}
\author{Masatoshi Imada}
\inst{Waseda Research Institute for Science and Engineering, Waseda University, Okubo, Shinjuku, Tokyo, 169-8555, Japan. \\
Toyota Physical and Chemical Research Institute, Yokomichi, Nagakute, Aichi, 480-1192, Japan.} 
\abst{Studies on the interplay between the charge order and the $d$-wave superconductivity in the copper-oxide high $T_{\rm c}$ superconductors are reviewed with a special emphasis on the exploration based on the unconventional concept of the electron fractionalization and its consequences supported by solutions of high-accuracy quantum many-body solvers. Severe competitions between the superconducting states and the charge inhomogeneity including the charge/spin striped states revealed by the quantum many-body solvers are first addressed for the Hubbard models and then for the {\it ab initio} Hamiltonians of the cuprates derived without adjustable parameters to represent the low-energy physics of the cuprates. The charge inhomogeneity and superconductivity are born out of the same mother, namely, the carrier attraction arising from the strong Coulomb repulsion near the Mott insulator (Mottness) and accompanied electron fractionalization. The same mother makes the severe competition of the two brothers inevitable.  The electron fractionalization has remarkable consequences on the mechanism of the superconductivity. Recent explorations motivated by the concept of the fractionalization and their consequences on experimental observations in energy-momentum resolved spectroscopic measurements including the angle resolved photoemission spectroscopy (ARPES) and the resonant inelastic X-ray spectroscopy (RIXS)  are overviewed, with future vision for the integrated spectroscopy to challenge the long-standing difficulties in  the cuprates as well as in other strongly correlated matter in general. 
}

\begin{document}
\maketitle

%
\section{Introduction}
\label{Sec.1}

Electronic charge inhomogeneity driven by Coulombic mutual electron repulsion 
often makes the spontaneous translational symmetry breaking accompanied by electron rich and poor regions in real space and is widely observed in strongly correlated electron systems.
If the inhomogeneity has a nonperiodic structure, it shares a conceptual similarity with Anderson localization, caused by the extrinsic disorder such as atomic impurity potential. 
The nonperiodic charge inhomogeneity reminds us mesoscale static patch structure rather than that in the atomic scale.
Instead, the charge order is usually defined as the periodic structure of electronic density modulation whose period is longer than the unit cell length, 
in which the pure electronic origin due to the electron Coulomb interaction as well as the electron-phonon coupling such as cases leading to the Peierls transition are included as the driving force.

If the electrons are isotropically repelling each other by the long-range Coulomb interaction as in the case of the electron gas, 
the charge order may appear in the form of the Wigner crystal~\cite{Wigner1934PhysRev.46.1002}, where each single electron resides separately by keeping the mutual distance as long as possible. In this case the distance to the closest electron determines the period of the charge order. In 2D systems, it normally shows up as the triangular or hexagonal lattice structures. However, if the lattice structure of underlying atoms introduces anisotropies, and if the electron kinetic energy also introduces the anisotropy due to the band structure, the spatial structure of crystallized electrons becomes highly nontrivial. The period of the crystallization is still essentially scaled by the distance to the averaged nearest neighbor electrons or the inverse of the Fermi momentum $k_{\rm F}$.

It is important to mention that the crystallization of electrons becomes greatly stabilized when the periodic order satisfies the commensurability condition in the presence of the underlying lattice structure (namely, if a simple fractional ratio or multiplicity is satisfied between the electron periodicity and the underlying lattice period, or in other words, the band filling is a simple fractional number or integer)~\cite{NodaImada2002PhysRevLett.89.176803} in comparison to the uniform continuum~\cite{ImadaTakahashiWigner1986,TanatarCeperley1989PhysRevB.39.5005}. In the absence of the underlying lattice and the commensurability condition, the charge order would not be stabilized in the realistic carrier density of the cuprates and the order would be limited to the region with orders of magnitude smaller carrier densities. The Mott insulator itself is the most typical case of the stabilized crystal with the strong commensurability, where the electron filling is an odd integer. If the original band filling satisfies the even integer, of course, it can be interpreted as the band insulator without the charge ordering. 
Since the superconductivity is not particularly stabilized by the commensurability condition, but the charge order is, it is crucially important to understand this circumstance in analyzing their competitions.     

The segregation of electrons and the resultant inhomogeneity  is not only triggered by the repulsion but also by the effective attractive force as is well known in the classical examples of atomic crystallization and phase separation (PS).
The long-ranged Coulomb repulsion prohibits the macroscopic PS if the underlying atom positions are fixed,
while the electrons are repulsively interacting at short enough distances without screening. However, in the intermediate range of distances and associated time retardation, the effective attraction may appear.

\begin{figure}[h]
  \begin{center}
   \includegraphics[width=50mm]{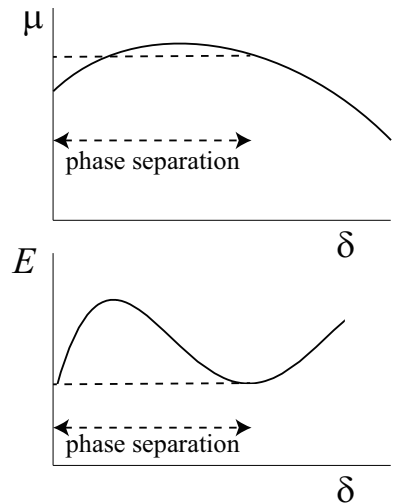}
  \caption{(Color online) Schematic features of energy $E$ and chemical potential $\mu$ vs. the doping concentration $\delta$ for doped Mott insulator. The Maxwell rule determines the phase separation (PS) region from the contact point of the tangential line with the energy curve $E$ in the lower panel.  
  }
  \label{PhaseSeparation}
\end{center}
\end{figure}

Moreover, the original Coulomb repulsion generates emergent attraction~\cite{Misawa2014a,ImadaSuzuki}, when the underlying Mott insulator affects the interaction of dilute carriers resulted from the carrier doping into the Mott insulator.  
The attraction is directly caused by the nonlinear reduction of the kinetic energy as a function of the doping concentration and thereby the total energy behaves similarly as is illustrated in Fig.~\ref{PhaseSeparation}. With increasing the doping concentration $\delta$ measured from the Mott insulator, the nonlinear reduction is the consequence of the fact that the carrier rapidly gets its mobility and recovers the coherence which are prohibited in the Mott insulator. The nonlinear reduction of the energy is represented as $E\propto -\frac{1}{2}b \delta^2$ with $b>0$.  The positive $b$ means nothing but the attraction of the carriers. This instantaneous attraction is a highly nontrivial consequence of the Mott physics because the original electron-electron interaction is strongly repulsive, which naively anticipates $b<0$.
In the cuprate high-$T_{\rm c}$ superconductors, the charge order, inhomogeneities  and fluctuations are likely to be primarily driven in such a circumstance, which is our focus in this article.   

The most extensively studied theoretical models for the charge order and fluctuations are the Hubbard model and its extensions especially on the two-dimensional lattice motivated by various stripe-type orders found near the superconducting phase of the cuprate superconductors~\cite{Tranquada1995,FradkinKivelsonTranquada2015} as well as in wider class of transition metal oxides~\cite{ImadaFujimoriTokura} and organic conductors~\cite{TakahashiNogamiYakushi,KanodaJPSJ2006,Hotta_Crystals_2012,Monceau2012,Clay_Mazumdar_2019,DresselTomic_2020}. The incentive of the extensive studies is coming from the observation that the Hubbard models seem to capture some of the essence of electron correlation effects in these compounds, despite its oversimplification in important details.  It has turned out that various periods of the charge order found in the Hubbard models have the energies extremely close each other and nearly degenerate also with the superconducting state as well as the charge uniform antiferromagnetic state consistently with the general experimental trend~\cite{Ido2018,Darmawan2018,Zheng2017}.  Though the energies are close, recent progress in the accuracy of numerical solvers for quantum many-body problems has enabled to reliably determine the phase diagram of the ground state of the Hubbard models including the charge order in the parameter space of the correlation strength, and the electron density. 


Since various different orders are severely competing each other even in real materials, it is required to study them by the first principles method without adjustable parameters to understand physics of the competition quantitatively and realistically. The most severe competitions are found between the superconductivity and various charge ordered states, both of which are the consequences of the effective attractive interaction. Since the conventional density functional theory does not have enough accuracy in the strongly correlated electron systems, accurate first principles framework suited for the correlated electron systems is imperative. Such a method has been developed recently as the multi-scale {\it ab initio} scheme for correlated electrons (MACE)~\cite{Imada_Miyake2010}. 

The feature of charge excitations and fluctuations have also been studied experimentally using recently developed spectroscopic methods such as the ARPES~\cite{Damascelli2003}, scanning tunneling microscope (STM)~\cite{Pan2001}, momentum resolved electron energy loss spectroscopy~\cite{Abbamonte2017} and X-ray scattering for the high-$T_{\rm c}$ superconductors including the cuprates.  Microscopic probes such as nuclear quadrupolar resonance have also been widely studied to understand the charge inhomogeneity and disproportionation.  We focus on charge order and fluctuations revealed by these studies in collaboration with theoretical analyses.   

The purpose of this review is to supply an overview of recent studies performed to understand the competitions of the charge order and superconductivity as well as the charge excitations in the cuprate high-$T_{\rm c}$ superconductors, which also provides us with insights into the mechanism of the superconductivity at a deeper level.   
In Sect.~ 2, we review studies on the phase diagram associated with the electronic inhomogeneity and superconductivity for simple lattice models on two dimensional lattices.
The competition between the charge order and the superconductivity is highlighted and the mechanism and consequences of the competition are discussed.
In Sect.~3, we summarize how the insights gained for the theoretical models are realistically working or not in the {\it ab initio} Hamiltonians of cuprate and iron-based superconductors. 
Relevance of the concept of electron fractionalization becomes clarified from the results of Sects.\ref{Sec.2} and \ref{Sec.3}. In Sect.~\ref{Sec.4}, studies on consequences of the electron fractionalization supported in Sects.~\ref{Sec.2} and \ref{Sec.3} are reviewed in connection to the nature of the single-particle Green's function measured by ARPES.  In Sect.~\ref{Sec.5}, consequences of the fractionalization are further elucidated in relation to the two-particle spectroscopic quantities to be measured in RIXS. 
Section \ref{Sec.6} is devoted to summary and outlook.

\section{Charge fluctuations and order in Hubbard models}
\label{Sec.2}
\subsection{Models and physical quantities to be analyzed}
\label{Sec.2.1}
Despite its simple form of the Hubbard Hamiltonian, its ground state on the 2D square lattice in the parameter space of the onsite interaction $U$, and carrier concentration $\delta$ requires a highly accurate solver because of severe competitions of the charge order with various periods of stripe structure, the $d$-wave superconducting order and the antiferromagnetic order.  
The Hubbard model reads 
\begin{eqnarray}
  \mathcal{H} =&& -\sum_{i,j,\sigma} t_{ij}c^\dagger_{i\sigma} c_{j\sigma}+ U\sum_i^{N_s} n_{i\uparrow}n_{i\downarrow},
  \label{Hubbard}
\end{eqnarray}
where we restrict nonzero hopping amplitude $t_{ij}$ only to the nearest neighbor pair $t_{ij}=t$, and the next nearest neighbor pair $t_{ij}=t'$.
$U$ is the onsite repulsive interaction, $N_s = L \times L$ is the square lattice size, $c^\dagger_{i\sigma}$ ($c_{i\sigma}$) is a creation (annihilation) operator of an electron with spin $\sigma$ on the site $i$, and $n_{i\sigma} = c^\dagger_{i\sigma}c_{i\sigma}$. 
Hereafter, we take the lattice constant as the length unit. The hole carrier concentration measured from half filling is denoted by $\delta=1-N_e/N_s$, where $N_e$ is the electron number. 

Severe competitions are elucidated in this model by using various methods with the help of recently developed highly accurate solvers such as auxiliary-field quantum Monte Carlo, density matrix renormalization group, density matrix embedding~\cite{Zheng2017}, variational Monte Carlo~\cite{Ido2018,Darmawan2018}, and tensor network~\cite{Zhao2017,Corboz2014}. The results are now essentially consistent each other as we summarize here.    


To understand the nature of the solution, several physical quantities are useful to characterize the ground state.
To clarify antiferromagnetic order and fluctuation, the equal-time spin structure factor 
\begin{eqnarray}
S_{\rm S}(\bm{q})=\frac{1}{3N_s} \sum_{i,j} \braket{\bm{S}_i \cdot \bm{S}_j}e^{-i\bm{q}\cdot(\bm{r}_i-\bm{r}_j)}
\label{Sq}
\end{eqnarray}
is useful while the charge order/fluctuation can be examined by  
the equal-time charge structure factor 
\begin{eqnarray}
S_{\rm C}(\bm{q})=\frac{1}{N_s} \sum_{i,j} \braket{n_i n_j-\langle n\rangle^2}e^{-i\bm{q}\cdot(\bm{r}_i-\bm{r}_j)},
\label{Nq}
\end{eqnarray}
where $\langle n \rangle=N_e/N_s$ is the averaged electron density. 
The possibility of PS can be detected from the carrier concentration dependence of the total electronic energy $E$. When $d^2E/d\delta^2<0$, the system is unstable with negative charge compressibility $\kappa$, because $\kappa^{-1}=-d\mu/d\delta=d^2E/d\delta^2$. The PS region is determined from the Maxwell rule in two equivalent fashions illustrated in Fig.~\ref{PhaseSeparation}. One way is to draw the tangential line to the energy curve $E(\delta)$ from the Mott insulating point as was schematically illustrated in the lower panel of Fig.~\ref{PhaseSeparation} and as will be quantitatively clarified in Fig.~\ref{Misawa2014Fig3} later.  

Superconducting order/correlation can be monitored by $d_{x^2-y^2}$-wave superconducting correlation functions 
\begin{eqnarray}
P_d(\bm{r})=\frac{1}{2N_s}\sum_i \braket{\Delta^\dagger_d(\bm{r})\Delta_d(\bm{r}+\bm{r}_i) + \Delta_d(\bm{r})\Delta^\dagger_d(\bm{r}+\bm{r}_i)}
\label{Pd}
\end{eqnarray}
with
\begin{eqnarray}
\Delta_d(\bm{r}_i)=\frac{1}{\sqrt{2}}\sum_{\bm{r}} g(\bm{r})(c_{\bm{r}_i\uparrow}c_{\bm{r}_i+\bm{r}\downarrow} - c_{\bm{r}_i\downarrow}c_{\bm{r}_i+\bm{r}\uparrow}).
\end{eqnarray}
The form factor $g(\bm{r})$ is defined as
\begin{eqnarray}
g(\bm{r})=\delta_{r_x,0}(\delta_{r_y,1}+\delta_{r_y,-1})-\delta_{r_y,0}(\delta_{r_x,1}+\delta_{r_x,-1})
\end{eqnarray}
for the $d_{x^2-y^2}$ symmetry.
The long-range order of the spin and charge are conventionally estimated by 
$\lim_{L\rightarrow \infty} S(\bm{q})/N_{s}$ and  $\lim_{L\rightarrow \infty} N(\bm{q})/N_{s}$ 
for the momentum $\bm{q}$ at the peak to see the growth to the Bragg point.
The spin/charge order parameters are defined as $\Delta_{s/c}=\sqrt{S_{\rm S/C}(\bm{q}_{\rm peak})/N_s}$.
For the superconductivity, the long-range order in the thermodynamic limit can be defined by 
\begin{eqnarray}
\bar{P_d}=\frac{1}{M}\sum_{\sqrt{2}L/4<|\bm{r}|\leq\sqrt{2}L/2} P_d(\bm{r}),
\end{eqnarray}
where $M$ is the number of vectors satisfying $\sqrt{2}L/4<|\bm{r}|\leq\sqrt{2}L/2$ to exclude the short-ranged non-asymptotic part and the part affected by the boundary effect.
The superconducting order parameter is defined as $\Delta_{\rm SC}=\sqrt{\bar{P_d}}$.

%
%
%
\subsection{Phase diagram of $t$-$t'$ Hubbard model}
\label{Sec.2.2}
%
\begin{figure}[h]
  \begin{center}
 \includegraphics[width=75mm]{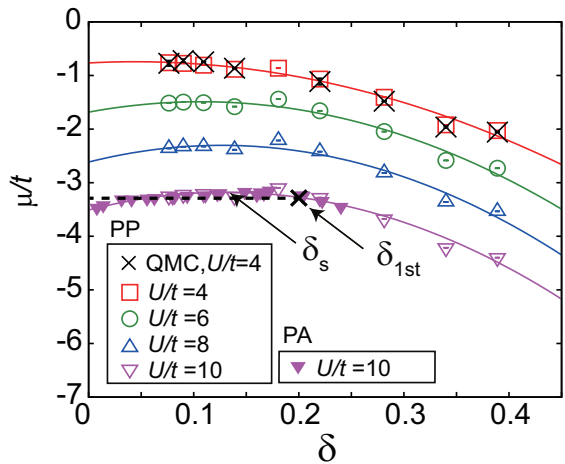}
  \caption{(Color online)
Doping concentration dependence of chemical potential for $U/t=4,8,6,10$ for the Hubbard model (reproduced from Misawa {\it et al.} (Ref.~\citen{Misawa2014a}, (c) [2014] American Physical Society.)).
Here, the system size dependence is negligible.
The PS region ($\delta<\delta_{\rm 1st}$) can be estimated
by {performing Maxwell's construction using the fitted second-order polynomials.}
{
As an example, Maxwell's construction for $U/t=10$ is shown by black dashed line.}
$\delta_s$ determined from the maximum of $\mu$ indicates the spinodal point where $\kappa$ diverges and the system becomes unstable for $\delta\leq \delta_s$.  }
\label{Misawa2014Fig3}
 \end{center}
\end{figure} 
The question about the electronic PS has long been a controversial issue for the Hubbard model.
In the initial stage, the PS was proposed based on small-size exact diagonalization study for the $t$-$J$ model derivable in the strong coupling limit~\cite{Emery1990}, while the quantum Monte Carlo study in the intermediate coupling, $U/t=4$ has suggested that the system is at criticality where the charge compressibility $\kappa$ diverges only in the Mott insulator limit, namely, $\lim_{\delta\rightarrow 0}\kappa\rightarrow \infty$, meaning that the system is at the quantum critical point of the PS at $\delta=0$~\cite{Furukawa1991,Furukawa1992}. Although the existence of the PS region for large $J/t$ in the $t$-$J$ model is rather trivial because of the explicit attraction introduced by the superexchange $J$, the issue is highly nontrivial for the Hubbard model.  The numerical methods were extensively applied to this issue, but the controversy remained~\cite{Dagotto1992PhysRevB.45.10741,Capone2006,Aichhorn2007,Khatami2010,Chang2010,Sordi2012,Neuscamman2012,Yokoyama2013,Tocchio2013,Misawa2014a,Sorella2021} (see also Table I of Ref.\citen{Misawa2014a} for the list of other comparisons).

\begin{figure}[h]
  \begin{center}
    \includegraphics[width=6cm]{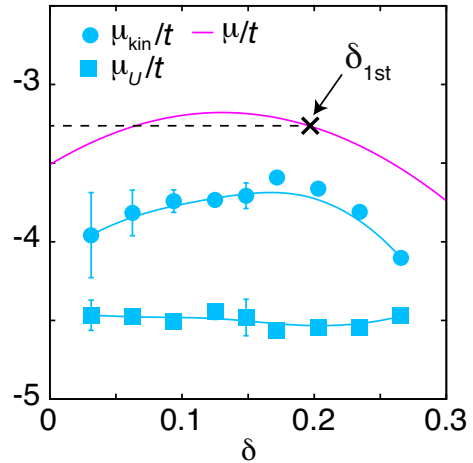}
 \caption{(Color online)
Doping concentration dependence of kinetic (potential) part of
chemical potential $\mu_{\rm kin}$ ($\mu_{U}$)
for $U/t=10$.
Solid curves are guides for eyes.
The total chemical potential $\mu$ is the same  as that shown  for $U/t=10$ in Fig.~\ref{Misawa2014Fig3}.
Black dashed line represents the line 
that is used for Maxwell's construction.
For better visibility, $\mu_{\rm kin}$ 
is shifted by {$-U/2$}. (Reproduced from Misawa {\it et al.} (Ref.~\citen{Misawa2014a}, (c) [2014] American Physical Society.))
  }
  \label{MisawaImada2014Fig14}
\end{center}
\end{figure}

Recently, the original controversy was revisited and it has come to the consensus that the diverging and negative charge compressibility with instability to the charge inhomogeneity appears in an extended region of the doping concentration for stronger coupling region $U/t>4$, if the charge homogeneity is imposed as one sees in 
Fig.~\ref{Misawa2014Fig3} for the simplest Hubbard model with only the nearest neighbor hopping and the onsite interaction $U$. This has also solved the initial controversy between the $t$-$J$ model and the Hubbard model. The PS indicated by the concave downward curve of the chemical potential appear if $U/t$ exceeds 4 and the Maxwell construction indicates that the PS region becomes wider and it is as large as $0<\delta\lesssim 0.2$ for $U/t=10$.
From Fig.~\ref{Misawa2014Fig3}, the chemical potential is estimated to be $\mu/t=a+b\delta+c\delta^2$, with $a=-3.51, b=5.09$ and $c=-19.51$ resulting in $E=E_0-a\delta-\frac{1}{2}b\delta^2-\frac{1}{3}c\delta^3$ at $U/t=10$. Therefore, the attractive interaction $v_{\rm att}(r_i-r_j)$ between two carriers integrated over space is estimated as $\sum_i v_{\rm att}(r_i)=5.09t$, which is surprisingly large. 
  The origin of the instability was identified as the $\delta$ dependence of the kinetic energy $E_{\rm kin}$ rather than the interaction energy as is illustrated in Fig.~\ref{MisawaImada2014Fig14}.

\begin{figure}[htbp]
  \begin{center}
   \includegraphics[width=7cm]{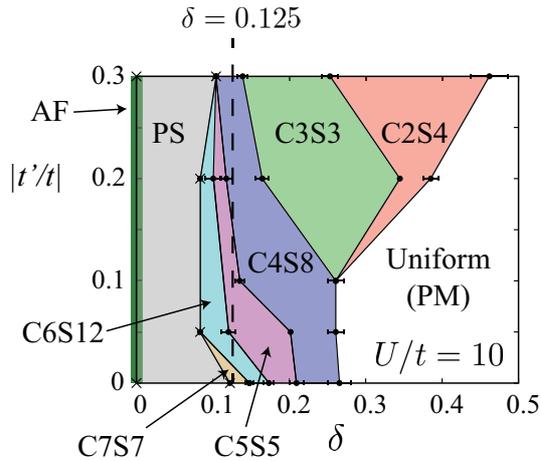}
  \caption{(Color online)
Ground-state phase diagram of the Hubbard model on a square lattice for $U/t=10$ reproduced from Ido {\it et al.} (Ref.~\citen{Ido2018}, (c) [2018] American Physical Society). Note that $t'/t$ is negative.
    At $\delta=0$, the ground state is the antiferromagnetic (AF) Mott insulator (green bold line). 
    Cross symbols indicate the calculated boundary of the PS and the stripe states.
   Solid black circles represent the calculated boundaries of ``C$l_c$S$l_s$" stripe states with $l_c$/$l_s$ period for charge/spin.
    Dashed line shows $\delta=0.125$.
    Solid lines and painted regions are guides for the eyes.
    In the unpainted (white) region, the ground state is a charge uniform paramagnetic (PM) state.
  }
  \label{pd}
\end{center}
\end{figure}

When the microscopic charge inhomogeneity characterized by the charge order with the periodicity longer than the lattice constant is allowed, recent high-accuracy simulations have revealed that the substantial part of the PS region is replaced by the stripe ordered phases with various periods.   
In Figure~\ref{pd}, the phase diagram of the Hubbard model is shown in the parameter space of $t'/t$ and $\delta$ with nonzero $t$ and $t'$ in the strong coupling region $U/t=10$ obtained by the variational Monte Carlo method~\cite{Ido2018}. Other methods give essentially the same tendency of competitions~\cite{Ponsioen2019}.
In Fig.~\ref{pd}, the stripe state with charge (spin) period $l_c$($l_s$) is denoted as ``C$l_c$S$l_s$''. 
As shown in Fig. \ref{pd},  charge inhomogeneous states exist as the ground states in a wide range of $\delta$ for any $t'/t$.
In the low doping limited region $\delta<0.1$, the PS region is found sandwiched by the antiferromagnetic insulator and stripe phases.
In the region $-0.3 \leq t'/t \lesssim -0.15$, which is a realistic range of $t'/t$ for the cuprates, the ground state at $\delta=1/8$ is the C4S8 state which has been observed in {\rm La}-based cuprates \cite{Tranquada1995,Tranquada1996}.
Since the macroscopic PS is not allowed in realistic materials with the long-ranged Coulomb interaction, the PS replaced by the stripe order in the majority of doping region happens to alleviate the unrealistic aspect of the Hubbard model with only short-ranged interaction.  
The agreement of the stripe period with the experimental indications at realistic $|t'/t|\sim 0.2$ also implies that the Hubbard model captures some realistic aspects of the cuprates. However, the relation to the superconductivity challenges the simple Hubbard model in terms of the realistic Hamiltonian of the cuprates as we discuss below.
%
%

\begin{figure}[htbp]
  \begin{center}
   \includegraphics[width=7.5cm]{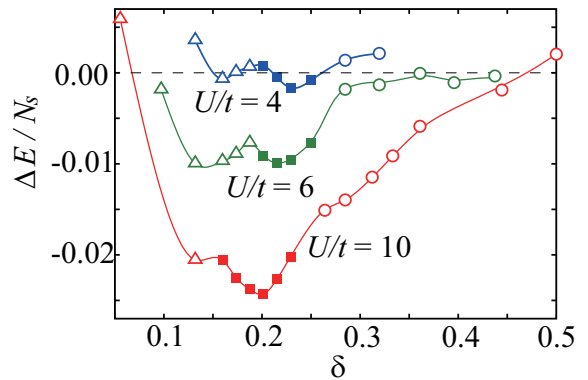}
 \caption{(Color online)
Interaction dependence of the stability of uniform and inhomogeneous states (the energy difference $\Delta E = E_{\rm stripe}-E_{\rm uniform}$) for $t'/t=-0.3$ reproduced from Ido {\it et al.} (Ref.~\citen{Ido2018}, (c) [2018] American Physical Society). 
Here, $E_{\rm stripe}$ and $E_{\rm uniform}$ are the energies of stripe and uniform states, respectively, in the unit of $t$. 
Circle, square and triangle symbols show the energies of  C2S4, C3S3, and C4S8 stripes, respectively. 
Red (in bottom), green (in middle) and blue (in top) symbols represent $\Delta E$ for $U/t=10$, $6$, and $4$, respectively. 
Curves are guides for the eyes.}
\label{diff_tp-0.3}
\end{center}
\end{figure}

%
%
%
To understand the relation between the stability of the stripe order and the interaction, the interaction dependence of the energy difference between the spatially uniform (superconducting) and the stripe states for $t'/t=-0.3$ is shown in Fig. \ref{diff_tp-0.3}~\cite{Ido2018}. 
The stripe states emerge as the ground states above $U/t \sim 4$ and the stripe phase becomes widened with the increase in $U$ as one sees in Fig.~\ref{diff_tp-0.3}.
The stabilization of the stripe order for larger $U/t$ offers a consistent and comprehensive picture for its mechanism: The effective attractive interaction increases with $U$ as indicated by the amplitudes of the negative curvature of the curves in Fig.~\ref{Misawa2014Fig3}. 

\begin{figure}
\begin{center}
 \includegraphics[width=0.5\textwidth]{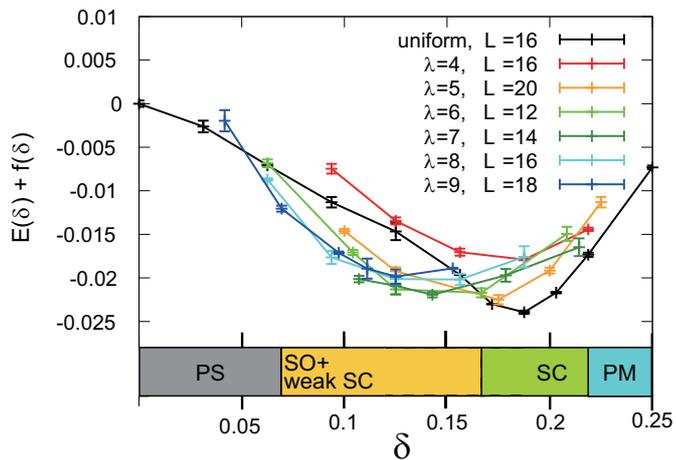}
   \caption{(Color online) Energies of stripe and uniform states {per site in units of $t$} as a function of doping for the length in $y$ direction $L_y=8$ systems ($U/t=10$) (Reproduced from Darmawan {\it et al.} (Ref.~\citen{Darmawan2018}, (c) [2018] American Physical Society)). The length in $x$ direction $L=L_x$ is indicated in the legend. As finite size effects are found to be small the result here essentially represents the thermodynamic limit.  
    A linear function $f(\delta)= 1.835\delta+0.4304$ has been added to the energy to improve visibility. (Lower strip panel): Ground state phase diagram of PS,
stripe spin and charge order coexisting with weak (vanishing) $d$-wave superconductivity (SO $+$ weak SC),
charge uniform $d$-wave superconductivity (SC) and charge uniform paramagnetic normal (PM) regions. The region $0.12<\delta<0.19$ is subject to the PS between the SO at $\delta \sim 0.12$ and SC at $\delta \sim 0.19$ as is speculated from the possible existence of the cotangent line to connect the ground states.}
\label{f:phase_diagram}
\end{center}
\end{figure}
The phase diagram was also examined at $t'/t=0$ and $U/t=10$ by using further improved variational wavefunction achieved by the combination with the tensor network~\cite{Zhao2017}.
The phase diagram is shown in Fig.~\ref{f:phase_diagram}~\cite{Darmawan2018}.
This also confirms that the energies of various periods of stripes and charge uniform $d$-wave superconducting state are nearly degenerate and the energy differences are around 0.01$t$ or less corresponding to $<0.005$ eV in the cuprate energy scale.  
After the extrapolation to the thermodynamic limit, the superconducting order seems to survive near $\delta \approx 0.2$. 
 Below $\delta \approx 0.19$, possible PS between charge uniform superconducting state and the stripe ordered state is observed, where the stripe order possibly coexists with very weak superconducting order. However, the coexistence of the stripe order with the very weak $d$-wave superconductivity is still a controversial issue, but recent highly accurate numerical results consistently show that a wide region roughly $0<\delta<0.2$ is governed essentially by the stripe ordered state~\cite{Zheng2017,Ido2018,Darmawan2018,Corboz2014,Ponsioen2019}. The dominance of the stripe state is not consistent with a wide region of the charge uniform superconducting phase universally observed in the cuprates.

\section{Competition between superconductivity and charge order in {\it ab initio} phase diagram of cuprate superconductors}
\label{Sec.3}
\begin{figure}[h!]
\begin{center}
  \includegraphics[width=8cm]{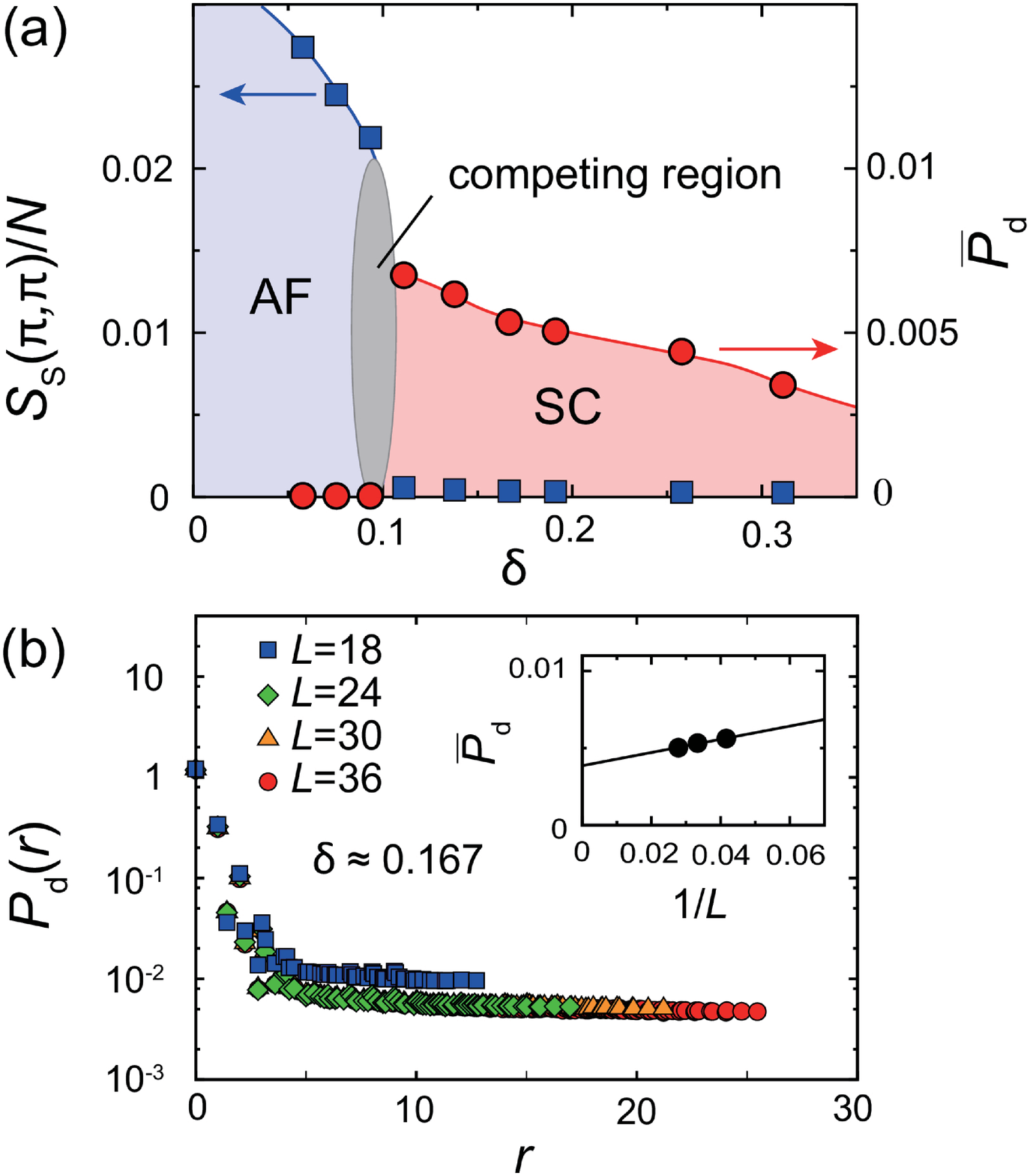}
\caption{ (Color online) (a) Physical quantities $S_{\rm S}(\pi, \pi)/N$ and ${\overline P}_{d}$ of homogeneous states for the system size $L=30$ as functions of $\delta$. Gray region indicates a region where antiferromagnetic (AF), superconducting (SC), and stripe states are severely competing. Energy competition between the stripe and superconducting states is shown in Fig. \ref{fig:stripes}. (b) Size dependence of $P_{d}(r)$ at $\delta \simeq 0.167$. In the inset, ${\overline P}_{d}$ ($L=24$, 30, and 36) is extrapolated to the thermodynamic limit.
(Reproduced from Ohgoe {\it et al.} (Ref.~\citen{Ohgoe2020}, (c) [2020] American Physical Society)).}
\label{fig:phys_uniform}
\end{center}
\end{figure}

\begin{figure}[h!]
\begin{center}
  \includegraphics[width=7cm]{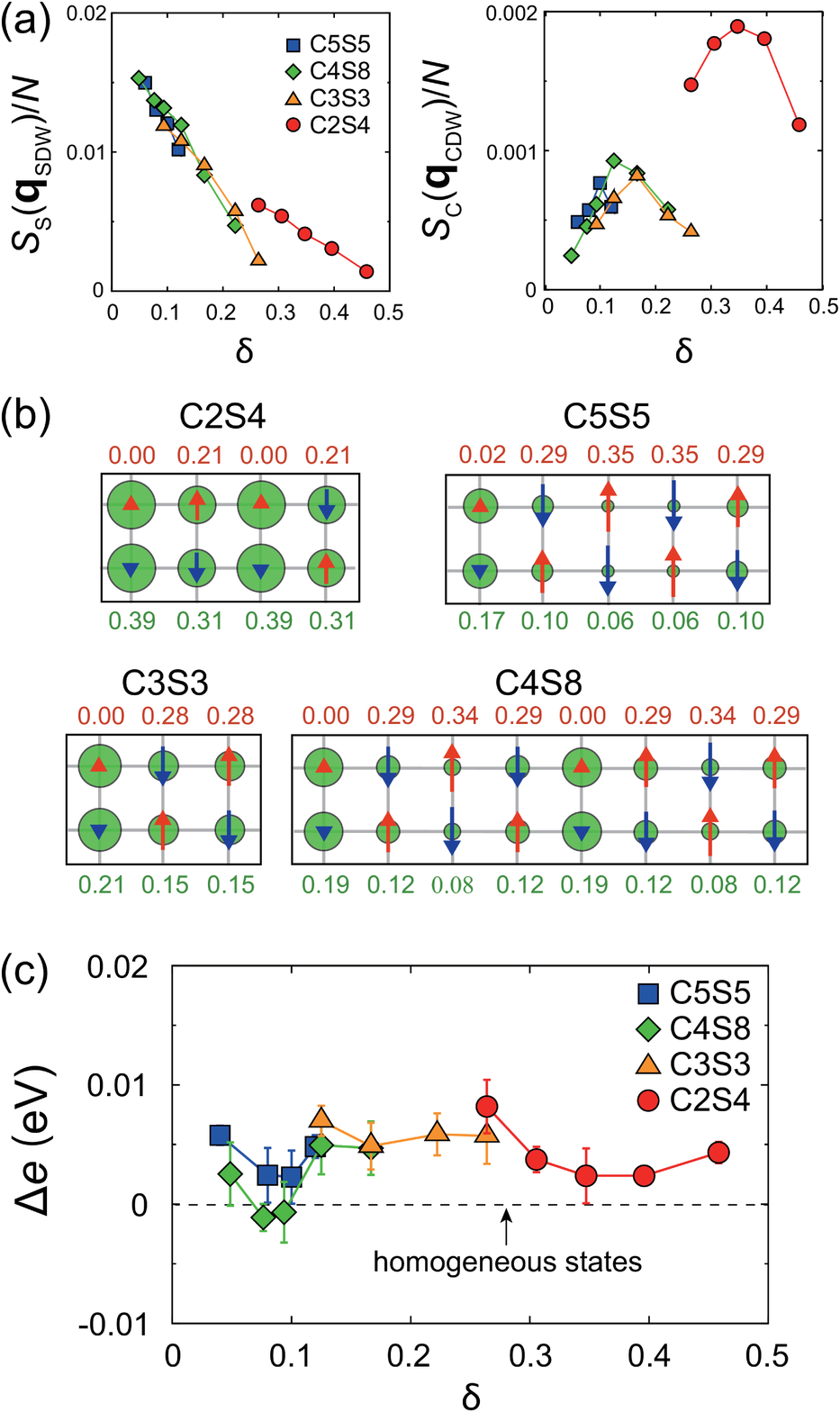}
\caption{ (Color online) Physical quantities of stripe states. (a) $S_{\rm S}({\bm q}_{\rm SDW})/N$ and $S_{\rm C}({\bm q}_{\rm CDW})/N$ of stripe states as functions of $\delta$. ${\bm q}_{\rm SDW}$ and ${\bm q}_{\rm CDW}$ are the momenta at the peak of $S_{\rm S}$ and $S_{\rm C}$, respectively. The linear system sizes are $L=20$ for C5S5 and $L=24$ for others. (b) Spin/charge configurations of several stripes. The hole concentrations are $\delta=0.1$ for C5S5, $\delta=0.125$ for C4S8, $\delta \simeq 0.167$ for C3S3 and $\delta \simeq -.347$ for C2S4. The hole density $\delta = 1 - \langle n_i \rangle $ and the local spin moments $\langle n_{i,\uparrow} -n_{i,\downarrow} \rangle/2$ are illustrated by the circle radius and the arrow length, respectively. Their values are also given as the green numbers and red numbers, respectively. (c) Stripe state energies relative to homogeneous states. (Reproduced from Ohgoe {\it et al.} (Ref.~\citen{Ohgoe2020}), (c) [2020] American Physical Society).}
\label{fig:stripes}
\end{center}
\end{figure}

To understand the experimental phase diagram quantitatively, and to elucidate the realistic mechanism of superconductivity in the families of the cuprates, the {\it ab initio} effective Hamiltonians of the cuprate superconductors, La$_{2-x}$Sr$_x$CuO$_4$ and HgBa$_2$CuO$_{4+y}$ were derived from the MACE procedure~\cite{Hirayama2018,Hirayama2019} and then they were  solved by the variational Monte Carlo method supplemented by the Lanczos and tensor network tools~\cite{Ohgoe2020}. The accuracy of the {\it ab initio} calculation is evidenced by the quantitative agreement of the Mott gap (2 eV) and the antiferromagnetic ordered moment (0.6 $\mu_{\rm B}$) with the experimental values of La$_2$CuO$_4$ without adjustable parameters, which has never been achieved before.  The phase diagram as a function of the carrier concentration $\delta$ estimated with the same accuracy is shown in Fig.~\ref{fig:phys_uniform} with the key physical quantity in each phase for HgBa$_2$CuO$_{4+y}$. The region of the experimental antiferromagnetic, stripe and $d$-wave superconducting ordered phases are quantitatively reproduced and the severe competition between the stripe and superconducting phases are quantitatively clarified as shown in Fig.~\ref{fig:stripes}. It has turned out that in a wide region the superconducting ground state is stabilized, while the stripe states with various periods exist as metastable excited states with the energies in the range of only 5 meV higher than the superconducting ground state.   

The  reproduced $d$-wave superconducting phase in the {\it ab initio} calculation provides us with  several insights. First, it was revealed that the strong effective attraction generated by originally the strong Coulomb repulsion caused by the Mott-physics mechanism simultaneously generates the two tendencies, one, the charge inhomogeneity and the other, the Cooper pairing. The severe competition between the stripe and the superconducting states is an intrinsic and inescapable property because they share the same roots of the effective attraction traced back to Mott physics. This means that to realize the higher $T_{\rm c}$ superconductivity, it is imperative to overcome this double-edged sword or the antinomy in the consequence of the effective attractive interaction.  

\begin{figure}[h]
\begin{center}
  \includegraphics[width=8cm]{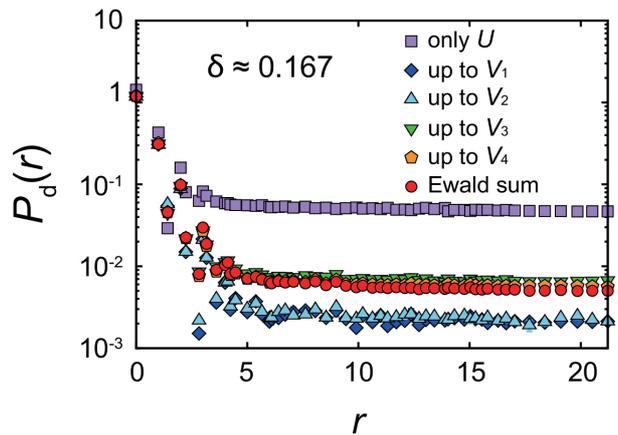}
\caption{(Color online) $P_d(r)$ at $\delta \simeq 0.167$ ($N_s=30\times 30$) for several choices of interaction ranges modified from the {\it ab initio} Hamiltonian of HgBa$_2$CuO$_{4+y}$. In the legends,``only $U$" represents that we truncated the off-site interactions and ``up to $V_n$" means that we included them up to $V_n$, where $V_n$ denotes the interaction with the $n$-th neighbor site. (Reproduction from Ohgoe {\it et al.} (Ref.~\citen{Ohgoe2020}), (c) [2020] American Physical Society).}
\label{fig:Vdep}
\end{center}
\end{figure}

In this respect, the role of the off-site Coulomb repulsion reported in Ref.~\citen{Ohgoe2020} is an interesting finding. Figure~\ref{fig:Vdep} shows that by switching off the specific range of the Coulomb repulsion from the {\it ab initio} Hamiltonian, the amplitude of the superconducting order as well as the stability of the charge order sensitively changes. Although the nearest-neighbor repulsion $V_1$ severely suppresses the superconductivity, inclusion of the third and fourth neighbor repulsions partially recovers the superconductivity presumably by the geometrical frustration that strongly suppresses the stripe order and inhomogeneity. Accurate ab initio parameters of the interaction and the transfer in an extended spatial range are important to quantitatively reproduce the experiments.

The second insight is its support for the emergent fractionalization. The strong attraction caused by the Mottness also indicates the underlying PS into two phases, one, the Mott insulator (or underdoped Mott insulator) and the other the conventional overdoped metal. Such an underlying bistability emerging concomitantly with the quantum entanglement in the system provides a platform of the electron fractionalization into two different types of fermions, which is the subject of the following section.

\section{Consequence of electron fractionalization in single-particle spectroscopic data}
\label{Sec.4}
Electrons with an underlying bistability may also dynamically fluctuate between the two bistable characters, instead of generating the static charge order (segregation) in real space, namely uniformly generate the dual character of the electrons dynamically fluctuating between the nature of the underdoped carrier ($d$ fermion) and $c$ fermion with the conventional metal properties which is stabilized at overdoping. Such a quantum tunneling is represented by the two-component fermion model~\cite{Sakai2016a,ImadaSuzuki} defined by 
 the Hamiltonian
\begin{eqnarray}
H&=&\sum_{k,\sigma}[ \epsilon_c (k)c_{k,\sigma}^{\dagger}c_{k,\sigma} +\epsilon_d (k)d_{k,\sigma}^{\dagger}d_{k,\sigma}  \nonumber 
\\
&+& \Lambda (k) (c_{k,\sigma}^{\dagger}d_{k,\sigma} +{\rm H.c.}).
\label{TCfermion} 
\end{eqnarray}
Instead of the real space segregation, this phenomenological Hamiltonian represents the quantum entanglement of the dual nature with electron {\it fractionalization} into $c$ and $d$. In the overdoped region, the electrons (or holes) behave essentially as unbound fermions with itinerant quasiparticle character represented by $c$ in Eq. (\ref{TCfermion}). On the other hand, in the Mott insulator, the electrons and holes are strongly bound as excitons with the binding energy scaled by the Mott gap. Even in the lightly doped region, the carriers may continue to bear the character of this bound exciton though the binding energy of the exciton should be substantially decreased because of the screening by other carriers. 

 Irrespective of the nature of the carriers representing the underdoped region around one of the bistable point, this emergent fermion is represented by $d$  in Eq.(\ref{TCfermion}).
It was shown that this phenomenological Hamiltonian successfully describes the pseudogap 
by the hybridization gap of $c$ and $d$ through the self-energy pole (at $\omega=\epsilon_d(k)$) of the $c$ component Green's function~\cite{YRZ,Sakai2016a}.

Moreover, by introducing the anomalous term generated by the superconducting mean field as
\begin{eqnarray}
H&=&\sum_{k,\sigma}[ \epsilon_c (k)c_{k,\sigma}^{\dagger}c_{k,\sigma} +\epsilon_d (k)d_{k,\sigma}^{\dagger}d_{k,\sigma}  \nonumber 
\\
&+& \Lambda (k) (c_{k,\sigma}^{\dagger}d_{k,\sigma} +{\rm H.c.})
\nonumber 
\\
&+&(\Delta_c(k) c_{k,\sigma}^{\dagger}c_{-k,-\sigma}^{\dagger}+\Delta_d(k) d_{k,\sigma}^{\dagger}d_{-k,-\sigma}^{\dagger} + {\rm H.c})], \nonumber \\
\label{TCfermionAnomalous} 
\end{eqnarray}
the self-energy of the $c$ component Green's function has an unusual character where the pole of the normal self-energy which generates the pseudogap  cancels the pole of the anomalous self-energy in the contribution to the single-particle Green's function. More concretely, the normal Green's function of $c$ in the superconducting phase is represented as 
\begin{eqnarray}
G_c(k,\omega)=\frac{1}{\omega-\epsilon_c(k)-\Sigma^{\rm nor}(k,\omega)-W(k,\omega)},
\label{TCMSCGkw} 
\end{eqnarray}
with 
\begin{eqnarray}
W(k,\omega)=\frac{\Sigma^{\rm ano}(k,\omega)^2}{\omega+\epsilon_c(k)+\Sigma^{\rm nor}(k,-\omega)^*},
\label{TCMSCWkw} 
\end{eqnarray}
where
\begin{eqnarray}
\Sigma^{\rm nor}(k,\omega)=\frac{\Lambda(k)^2(\omega+\epsilon_d(k))}{\omega^2-\epsilon_d(k)^2-\Delta_d(k)^2},
\label{TCMSCSCSNkw} 
\end{eqnarray}
and 
\begin{eqnarray}
\Sigma^{\rm ano}(k,\omega)=\Delta_c(k)+\frac{\Lambda(k)^2\Delta_d(k)}{\omega^2-\epsilon_d(k)^2-\Delta_d(k)^2}.
\label{TCMSCSCSAkw} 
\end{eqnarray}
The residue of this pole in $\Sigma^{\rm nor}$ at $\omega =\pm\sqrt{\epsilon_d(k)^2+\Delta_d(k)^2}$ cancels with that of $W$ at the same $\omega$, which results in the cancellation of the poles in the normal and anomalous contributions in $G_c$.
This cancellation accounts for the absence of the anomaly in the Green's function for the Hubbard model~\cite{Sakai2016a} and is consistent with the absence of prominent anomalies in the spectral function $A(k,\omega)=-\frac{1}{\pi}{\rm Im}G(k,\omega)$ in the ARPES data of the cuprates. 

\begin{figure}[h!]
\begin{center}
\includegraphics[width=0.3\textwidth]
{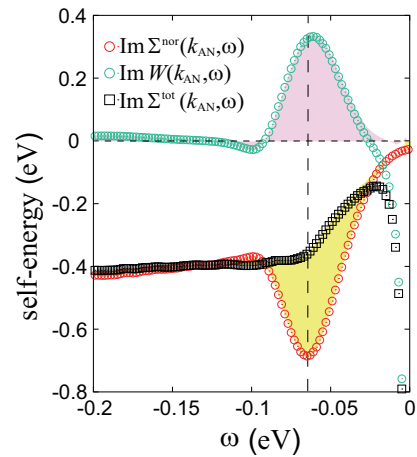}
\caption{(Color online)
Normal and anomalous self-energies derived from machine learning. (Reproduction from  Yamaji {\it et al. } (Ref.~\citen{Yamaji2021})).
The contribution of normal self-energy $\Sigma^{\rm nor}(k_{\rm AN},\omega)$ 
and anomalous self-energy ${\rm Im} W(k_{\rm AN},\omega)$ to the spectral function 
are separately deduced from the ARPES data from Kondo {\it et al.}~\cite{kondo2011disentangling}  
by the machine learning for  Bi2212 at the antinodal momentum $k_{\rm AN}$. The vertical dashed lines indicate the peak positions $\omega_{\rm PEAK}$ in the their peaks. 
The prominent peak structures
are identified as 
yellow shaded area for ${\rm Im} \Sigma^{\rm nor}(k_{\rm AN},\omega)$ 
 and 
pink shaded area for ${\rm Im} W(k_{\rm AN},\omega)$.
The yellow and pink areas cancel in their sum ${\rm Im} \Sigma^{\rm tot}(k_{\rm AN},\omega)$. 
}
\label{FigSigma}
\end{center}
\end{figure}

Figure~\ref{FigSigma}~\cite{Yamaji2021} has demonstrated that the Boltzmann machine learning of the ARPES data taken from a cuprate superconductor, Bi$_2$Sr$_2$CaCu$_2$O$_{8+\delta}$ (Bi2212)~\cite{kondo2011disentangling} 
extracts the emergent peaks in the contribution of the normal ($\Sigma^{\rm nor}$) and anomalous ($W$) components of the self-energies and reproduces their cancellation. 
This cancellation has turned out to be significant because it is hidden in the experiments while the peak in the anomalous self-energy directly generates the dominant part of the superconducting gap in the real part of the self-energy through the Kramers-Kronig relation.

In terms of the charge fluctuation, a possible interpretation is that $c$ may represent the carrier with the electronic negative charge while $d$ may represent the other bistable point possessing the property of the fermionic component of a weakly bound exciton consisting of an  electron bound to a hole that is charge neutral in total.  Therefore, a regular alignment of $c$ and $d$ generates the charge order with a charge excitation gap. The charge order in this mechanism becomes stabilized by the repulsive (or less attractive) force between $c$ and $d$ in comparison to the relatively stronger attraction between two $d$ fermions and/or between two $c$ fermions. The relative attraction of two $d$ particles may cause the pairing of the two as the origin of the anomalous term in Eq.~(\ref{TCfermionAnomalous}). On the other hand, the hybridization of $c$ and $d$ generates the pseudogap in the spectral function of $c$ and $d$ as is mentioned above.  The origin of attraction for $d$ can be associated with the kinetic mechanism arising from Mott physics as is discussed in Sec.\ref{Sec.2.2} evidenced in Figs.~\ref{Misawa2014Fig3} and \ref{MisawaImada2014Fig14}. The attraction may also be interpreted by the excitonic character, where the dipole moment associated with the exciton containing a $d$ fermion causes the dipole-dipole attraction including dispersive force. This attraction can be the order of the exciton binding energy or the intersite Coulomb interaction and is consistent in the energy scale with the attraction estimated by the quadratic dependence of the kinetic energy as is elucidated in the second paragraph of Sect.~\ref{Sec.2.2},  though the view is different.  

\section{Consequences of fractionalization in two-particle spectroscopic data} 
\label{Sec.5}
When we employ the phenomenological model of the fractionalization, the remarkable feature of the ARPES spectral function $A(k,\omega)$ was accounted for in Sect.~\ref{Sec.3}. Because this unconventional and distinct state of electrons has a large impact on the understanding of the mechanism of the cuprate superconductivity and must have effects in other experimental probes, it is desired to confirm possible experimental indications to further test the validity of the fractionalization and its effects on the charge correlation.
To reach the comprehensive understanding on the charge dynamics of the cuprates,  it would be powerful to perform integrated analyses by combining various spectroscopic methods, where each probe is able to detect only some projection of the whole momentum-energy resolved single- and two-particle dynamics. 
RIXS is one of spectroscopic probes which is able to reveal the two-body charge and spin correlations.   We here review a recent attempt of predicting the RIXS spectra to test the validity of the fractionalization.

In the RIXS measurement,  the X-ray excites a core electron to the conduction bands. A valence electron then decays to the core by X-ray emission leaving an exciton (electron-hole pair) near the Fermi level, which evolves in time until the pair recombination. The RIXS dynamics is expressed by the intensity of the emitted X-ray as a function of the momentum and energy transfers $Q$ and $\omega$, respectively, relative to the injected X-ray and the scattering intensity is given by   
\begin{eqnarray}
I_{\rm RIXS}(Q,\omega, \omega_0; \sigma, \rho)\propto\sum_l |B_{li}(Q, \omega_0;\sigma,\rho)|^2\delta(E_l-\omega), 
\label{I_RIXS_1}
\end{eqnarray}
\begin{eqnarray}
&&B_{li}(Q,\omega_0;\sigma,\rho)=\sum_{m,j}e^{iQ\cdot R_m}\chi_{\rho,\sigma}\langle l|c_{m\sigma}|j\rangle \nonumber
\\
&&\times\langle j|(\omega_0-E_j+i\Gamma)^{-1}|j\rangle\langle j |c^{\dagger}_{m\rho}|i(k=0)\rangle,
\label{I_RIXS_2} 
\end{eqnarray}
where 
$c_{m\sigma}^{\dagger}$ is the local electron creation operator at the $m$-th site with spin $\sigma$ while $\omega_0=E_j$ is the excitation energy of an eigenstate $|j\rangle$ of the Hamiltonian $H_m=H+V_m$ for the system with one electron excited from the core to the conduction band at the $m$-th lattice site by the X-ray injection to the initial ground state. The excitation energy $E_j$ of the conduction electron is measured  from the Fermi energy $E_{\rm F}$. The particle-hole (exciton) state $|l\rangle$ is characterized by the energy $E_l=\omega$ and the total momentum $Q$.  


The RIXS intensity contains in principle both spin and charge correlations, because if the core electron feels the spin-orbit interaction, the spin flipping transition is allowed in the process of the X-ray absorption and reemission.
Furthermore, if the core hole lifetime $1/\Gamma$ is long as in the case of O $K$ edge spectra, it contains richer information including electron and hole excitations separately in contrast to the simple spin or charge structure factor where they are involved as an instantaneous process. The effect of finite core hole lifetime was pointed out by Tohyama and Tsutsui~\cite{Tohyama2018}. 

\begin{figure}[h!]
\begin{center}
\includegraphics[width=0.45\textwidth]{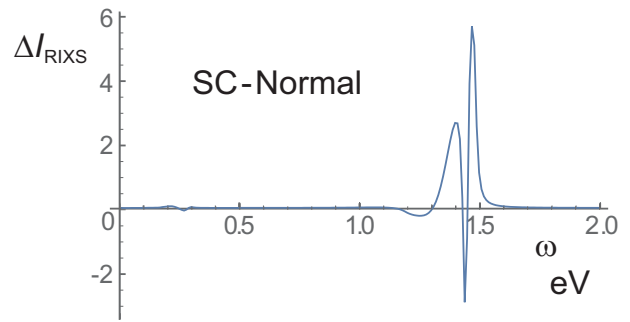}
\caption{(Color online)
Enhancement of RIXS intensity $\Delta_{\rm RIXS}(Q,\omega)$ defined by the difference between the intensity in  the superconducting phase and the normal (pseudogap) phase predicted from the fractionalized electron model (Eqs.~(\ref{TCfermion}) and (\ref{TCfermionAnomalous})) at $Q=(\pi,0)$ and at around the energy $\omega$ on the top of the ingap band determined from the TCFM parameters and also as roughly inferred from Fig.~\ref{Charlebois2020Fig6}.  (Reproduction from Ref.~\citen{ImadaRIXS2021}, (c) [2021] The Physical Society of Japan.)
\label{ImadaRIXS2021Fig12}
}
\label{FigAkw}
\end{center}
\end{figure}

\begin{figure}[h!]
\begin{center}
\includegraphics[width=0.3\textwidth]{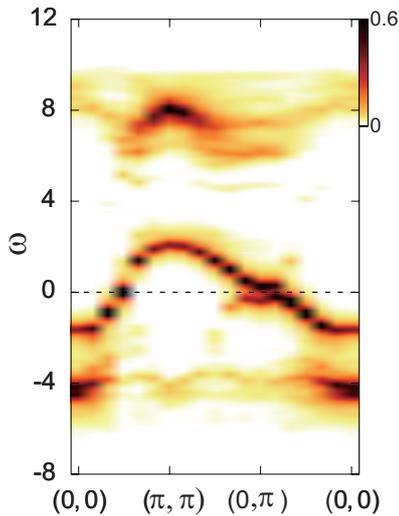}
\caption{(Color online) 
Momentum and energy dependences of  spectral function $A(k,\omega)=-\frac{1}{\pi}{\rm Im}G(k,\omega)$ of the Hubbard model with only the nearest neighbor term for the transfer at $U=8$ and $t=1$.
The spectral weight in the ingap band above the Fermi level $E_{\rm F}=0$ has a peak at $k=(\pi,\pi)$ and $\omega\sim 2t$, which corresponds to $\omega\sim 1$ eV, because $t$ in the {\it ab initio} Hamiltonian of the cuprates has the value $t\sim 0.5$ eV, while $U\sim 4$ eV is a realistic value~\cite{Hirayama2019}. Complexity arising from other off-site transfers and interactions do not affect this rough estimate of the peak energy at $k=(\pi,\pi)$.  (Reproduction from Charlebois {\it et al.} (Ref.~\citen{Charlebois2020}, (c) [2020] American Physical Society).)
\label{Charlebois2020Fig6}
}
\end{center}
\label{FigAkw}
\end{figure}

In the spin non flipping process, the fractionalized electron model (Eqs.(\ref{TCfermion}) and (\ref{TCfermionAnomalous})) was analyzed and it was predicted that the RIXS intensity is enhanced in the superconducting phase than the normal phase for specific energies and momenta if the electron is fractionalized~\cite{ImadaRIXS2021} as reproduced in  Fig.~\ref{ImadaRIXS2021Fig12} as an example. The enhancement is most prominently seen for the momentum transfer around $(\pm\pi,0)$ and $(0,\pm\pi)$, which contains the excitation of an electron at $(\pi,\pi)$ at the top of the ingap band (with the energy as large as 1 eV as demonstrated in Refs.~\citen{Kohno2012,Charlebois2020} and shown in Fig.~\ref{Charlebois2020Fig6}) and a hole at $(\pm\pi,0)$ and $(0,\pm\pi)$ (namely, decay of a valence electron at the antinode point near the Fermi level to the core hole).  Since in the conventional electron system the RIXS intensity is transferred from high to low energies in the superconducting state in comparison to the normal state, the enhancement at the energy as high as 1eV is a remarkable and unusual property of the fractionalized electron. The origin of the enhancement is explained by the following process: In the normal and pseudogap phase, a large portion of the electron is transferred to the $d$ component near the antinodal region, while the transition to the core hole represented by the conventional electron component is allowed only by the $c$ component, resulting in the suppression of the RIXS intensity. In the superconducting phase, the $c$ component increases due to the recovery from the strong hybridization, leading to the enhanced transition to the core hole.    

\section{Summary and Outlook}
\label{Sec.6}
It has been shown both in the Hubbard models and in the {\it ab initio} Hamiltonians of the cuprates without adjustable parameters that the carrier doping into the Mott insulators causes strong tendencies toward the charge order (stripe order) and the $d$-wave superconductivity with their severe competitions. 
Detailed studies of the {\it ab initio} and Hubbard Hamiltonians have both revealed that the following physics and process are at work.
The charge order (inhomogeneity) or its fluctuation ubiquitously observed there is understood as originated from the underlying effectively strong attractive interaction directly generated by the progressive but rapid release of the carriers upon carrier doping from the Mott localization. The nonlinear decrease (energy gain) in the kinetic energy with the evolution of the carrier doping leads to the negative curvature of the energy as a function of the doping, where the emergent attraction is the direct consequence of such a concave downward dependence of the kinetic energy induced by the release. The instability toward the charge inhomogeneity occurs because of this negative curvature. It causes the bistability of electrons and subsequent electron fractionalization into $c$ fermions representing the normal metal carrier more stable in the overdoped region, and $d$ fermions representing the barely localized carriers caused by the electron-hole binding that is more stable near the Mott insulator. The quantum tunneling between the two bistable states generates dynamical fluctuations, leading to the opening of the hybridization gap as the pseudogap. 

When the charge order is stabilized, it is realized by the alternating $c$ rich and $d$ rich regions, which is realized by the relatively stronger repulsive interaction between $c$ and $d$ fermions than the $c$ themselves and $d$ themselves.  Indeed the mutual attraction between the $d$ component is understood from the above release from the Mott localization as well as from the attractive nature of the dipole-dipole interaction between $d$, which possibly has the excitonic character.

Meanwhile, the attractive interaction also induces the instability to the Cooper pair formation, which is more prominent for the state at the underdoped part of the bistability represented by the $d$ component.  The itinerant charge component $c$ is also driven to the Cooper pair formation because of the hybridization with $d$ and hence coherent superconducting phase emerges. Note that the pair of $d$ fermions stays incoherent and charge neutral, which does not contribute to the phase coherent superconductivity. 

The model and {\it ab initio} studies both suggest that the charge order (spatial inhomogeneity) prevails the superconductivity if the carrier attraction is too strong (namely, the original Coulomb repulsion is too strong). The superconducting phase is stabilized in the intermediate coupling region. To enhance the stability of the superconductivity, it is useful to control the off-site interaction. Although the off-site repulsion suppresses the superconducting order, it suppresses the charge order and inhomogeneity more strongly if the geometrical frustration is arranged in the range and strength of the nonlocal Coulomb interaction to kill the charge ordering.  It was also reported that the design of the interface helps to suppress the electronic inhomogeneity by the self-doping and optimize the superconductivity~\cite{Misawa2016}.  The nonequilibrium state introduced by the laser pumping was also reported to help the suppression of the charge ordering and taking over by the superconductivity, where the oscillatory pump prohibits the static charge ordering~\cite{Ido2017}. Since the main mechanism of the superconductivity and  the severe competition with the charge  inhomogeneity has been clarified, we now have a possibility of designing materials synthesis and tune the experimental condition to control the stability of superconductivity and charge inhomogeneity.
 
To promote the detailed and quantitative understanding, combined and integrated spectroscopic probes are helpful if utilized systematically. The data science approaches are useful to analyze different spectroscopic results in a comprehensive and systematic manner
by overcoming the limitation of each spectroscopic data which often suffer from experimental noise and restricted range of measurements. Insights from first principles calculation faithfully based on the experimental structure of the cuprates and other strongly correlated electron systems are also useful to reach the whole understanding on this novel concept and physics. Such integrated spectroscopy science holds the key for the future progress. 


{\bf Acknowledgements}
The author thanks useful discussions and collaborations with Youhei Yamaji, Shiro Sakai, Marcello Civelli, Takahiro Misawa, Motoaki Hirayama, Kota Ido, Takahiro Ohgoe, Hui-hai Zhao, Andrew Darmawan, Yusuke Nomura, Terumasa Tadano, Maxime Charlebois, Atsushi Fujimori and Di-Jing Huang.
This work was financially supported by Grant-in-Aids
for Scientific Research (JSPS KAKENHI) (No. 16H06345) from Ministry of
Education, Culture, Sports, Science and Technology (MEXT), Japan. This work
was also supported in part by the projects conducted
under MEXT Japan named as ``Priority Issue
on Post-K computer" and ``Program for Promoting Research
on the Supercomputer Fugaku" in the subproject, ``Basic Science
for Emergence and Functionality in Quantum Matter:
Innovative Strongly-Correlated Electron Science by
Integration of Fugaku and Frontier Experiments".  We
also thank the support by the RIKEN Advanced Institute for Computational
Science through the HPCI System Research project (hp190145, hp200132 and hp210163)
supported by MEXT.

\bibliographystyle{jpsj_mod}
\bibliography{82407_after_proof}
\end{document}